\documentclass[aps,prl,twocolumn,amsmath,amssymb,superscriptaddress,nofootinbib,showpacs]{revtex4-1}

\usepackage{aas_macros}
\usepackage{graphicx}
\usepackage{bm}
\usepackage{wasysym}
\usepackage{enumitem}

\usepackage[colorlinks=true]{hyperref}

\begin{document}

\title{Testing Lorentz Symmetry with Lunar Laser Ranging}

\author{A.~Bourgoin}
\affiliation{SYRTE, Observatoire de Paris, PSL Research University, CNRS, Sorbonne Universit\'es, UPMC Univ. Paris 06, LNE, 61 avenue de l'Observatoire, 75014 Paris, France}
\email{adrien.bourgoin@obspm.fr}

\author{A.~Hees}
\affiliation{Department of Physics and Astronomy, University of California, Los Angeles, California 90095, USA}
\email{ahees@astro.ucla.edu}

\author{S.~Bouquillon}
\affiliation{SYRTE, Observatoire de Paris, PSL Research University, CNRS, Sorbonne Universit\'es, UPMC Univ. Paris 06, LNE, 61 avenue de l'Observatoire, 75014 Paris, France}
\email{sebastien.bouquillon@obspm.fr}

\author{C.~Le~Poncin-Lafitte}
\affiliation{SYRTE, Observatoire de Paris, PSL Research University, CNRS, Sorbonne Universit\'es, UPMC Univ. Paris 06, LNE, 61 avenue de l'Observatoire, 75014 Paris, France}
\email{christophe.leponcin@obspm.fr}

\author{G.~Francou}
\affiliation{SYRTE, Observatoire de Paris, PSL Research University, CNRS, Sorbonne Universit\'es, UPMC Univ. Paris 06, LNE, 61 avenue de l'Observatoire, 75014 Paris, France}
\email{gerard.francou@obspm.fr}

\author{M.-C.~Angonin}
\affiliation{SYRTE, Observatoire de Paris, PSL Research University, CNRS, Sorbonne Universit\'es, UPMC Univ. Paris 06, LNE, 61 avenue de l'Observatoire, 75014 Paris, France}
\email{m-c.angonint@obspm.fr}

\date{\today}


\begin{abstract}
  Lorentz symmetry violations can be parametrized by an effective field theory framework that contains both general relativity and the standard model of particle physics called the standard-model extension (SME). We present new constraints on pure gravity SME coefficients obtained by analyzing lunar laser ranging (LLR) observations. We use a new numerical lunar ephemeris computed in the SME framework and we perform a LLR data analysis using a set of 20721 normal points covering the period of August, 1969 to December, 2013. We emphasize that linear combination of SME coefficients to which LLR data are sensitive and not the same as those fitted in previous postfit residuals analysis using LLR observations and based on theoretical grounds. We found no evidence for Lorentz violation at the level of $10^{-8}$ for $\bar{s}^{T\!X}$, $10^{-12}$ for $\bar{s}^{X\!Y}$ and $\bar{s}^{X\!Z}$, $10^{-11}$ for $\bar{s}^{X\!X}-\bar{s}^{Y\!Y}$ and $\bar{s}^{X\!X}+\bar{s}^{Y\!Y}-2\bar{s}^{Z\!Z}-4.5\bar{s}^{Y\!Z}$ and $10^{-9}$ for $\bar{s}^{T\!Y}+0.43\bar{s}^{T\!Z}$. We improve previous constraints on SME coefficient by a factor up to 5 and 800 compared to postfit residuals analysis of respectively binary pulsars and LLR observations.
\end{abstract}

\maketitle


\par Since its establishment in 1915 by Einstein, general relativity (GR) has survived one century of experimental and theoretical scrutiny; its foundations have been tested spanning all scales, from the Solar System to the edge of the early Universe. Those tests can only be described in an extended framework parametrizing deviations from GR. In the past decades, two frameworks were widely used in the literature at the scale of the Solar System, namely, the parametrized post-Newtonian (PPN) \cite{1993tegp.book.....W,*2014arXiv1409.7871W} and the fifth force formalisms \cite{1999snng.book.....F}. However, some motivations are given to look for deviations to GR in other frameworks (see e.g. \cite{2012CQGra..29w5027H} and references therein), for example the standard-model extension (SME) framework \cite{1997PhRvD..55.6760C,1998PhRvD..58k6002C,2004PhRvD..69j5009K}. This framework aims at systematically parametrizing any hypothetical violation of the Lorentz symmetry in all sectors of physics from particles physics to gravity. 

\par Lorentz invariance is one of the fundamental symmetries of relativity and has profound implications that extend from particle physics to GR. It states that the outcome of any local experiment is independent of the velocity and of the direction of the local freely falling frame in which the experiment is performed \cite{1993tegp.book.....W}. Considering the broad field of applicability of this symmetry, searches for Lorentz symmetry breaking provide a powerful test of fundamental physics. In addition, many scenarios in the literature expect some Lorentz violations like, e.g., string theory, loop quantum gravity and noncommutative fields theory \cite{2014RPPh...77f2901T,2005LRR.....8....5M}. In this context, Colladay and Kosteleck\'y have built an effective field theory making possible confrontations between fundamental theories and experiments called the SME.

\par Following from \cite{2004PhRvD..69j5009K,2006PhRvD..74d5001B}, a hypothetical breaking of the Lorentz symmetry in the gravitational sector naturally leads to an expansion at the level of the gravitational part of the action which is given in the minimal SME as
\begin{align}
  S_g=&\frac{1}{2\kappa}\!\int\!d^4x\sqrt{-g}\Big(R\!-\!uR\!+\!s^{\mu\nu}R_{\mu\nu}^T\!+\!t^{\alpha\beta\mu\nu}C_{\alpha\beta\mu\nu}\Big)\nonumber\\
  &+S'[s^{\mu\nu},t^{\alpha\beta\mu\nu},g_{\mu\nu}]\text{,}\label{eq_SME}
\end{align}
with $\kappa=8\pi{}Gc^{-4}$ being the Einstein constant, $c$ being the speed of light in a vacuum, $g$ being the determinant of the metric tensor $g_{\mu\nu}$, $R$ being the Ricci scalar, $R_{\mu\nu}^T$ being the trace free Ricci tensor, $C_{\alpha\beta\mu\nu}$ being the Weyl conformal tensor, $u$, $s^{\mu\nu}$ and $t^{\alpha\beta\mu\nu}$ being the Lorentz violating fields. $S'$ contains the dynamical terms governing the evolution of SME coefficients. Note that Eq. \eqref{eq_SME} only contains Lorentz violating terms of the gravitational sector of the SME. The matter and electromagnetic sectors of the SME are not discussed here since they are constrained mainly by laboratory experiments (see \cite{2011RvMP...83...11K} for a summary of all constraints, and e.g., \cite{2014LRR....17....4W} for a discussion of the relation between the electromagnetic sector of the SME and previous frameworks like the $c^2$ formalism).

\par From experimental evidence the violating fields have to be small quantities. Therefore, it is reasonable to work in the linearized gravity limit where the metric depends only on $\bar{u}$ and $\bar{s}^{\mu\nu}$ which are the vacuum expectation values of $u$ and $s^{\mu\nu}$ \cite{2006PhRvD..74d5001B}. The coefficient $\bar{u}$ is unobservable since it can be absorbed in a rescaling of the gravitational constant. As mentioned by \cite{2006PhRvD..74d5001B} the so obtained post-Newtonian metric differs from the PPN one.

\par Many studies aimed at constraining the pure gravity SME coefficients by searching for possible theoretical signatures in postfit residuals of measurements analyzed in pure GR. This procedure has been applied to many observations: atom interferometry \cite{2009PhRvD..80a6002C}, gravity probe B \cite{2013PhRvD..88j2001B}, binary pulsars \cite{2013CQGra..30p5019S,2014PhRvL.112k1103S,*2014PhRvD..90l2009S}, planetary ephemerides \cite{2012CQGra..29q5007I,2015PhRvD..92f4049H}, cosmic ray observations \cite{2015PhLB..749..551K}, gravitational wave detection \cite{2016PhLB..761....1K}, and even lunar laser ranging (LLR) \cite{2007PhRvL..99x1103B}. In addition, several suggestions have been made to further constrain the SME coefficients \cite{2015PhRvD..92f4049H} e.g. using the Cassini conjunction data \cite{2003Natur.425..374B} to constrain the $\bar s^{TT}$ coefficient or to use the LAGEOS/LARES data that have been successfully used to detect the frame-dragging effect \cite{2004Natur.431..958C,2016EPJC...76..120C}.

\par However, as mentioned in \cite{2016arXiv160401663L} a postfit search for SME signatures is not fully satisfactory. In that paper, the authors showed that the uncertainties obtained by a postfit analysis based on a GR data reduction can be underestimated by up to 2 orders of magnitude. This is mainly due to correlations between SME coefficients and others global parameters (e.g., masses, positions, and velocities) that are neglected in this kind of approach. Moreover, the analytical Lorentz violation signatures that are looked for in this kind of postfit residuals analysis are always a combination of natural frequencies appearing in the fundamental problem governing the evolution of the experiment. Consequently, after a fit in pure GR, signals at the natural frequencies are absorbed in the redefinition of initial conditions and physical parameters. Therefore, it could be problematic to look only for main analytical Lorentz violating signals in postfit residuals since it could have been absorbed in a redefinition of one or more physical parameters. Finally, in the case of LLR data analysis, the oscillating signatures derived in \cite{2006PhRvD..74d5001B} and used in \cite{2007PhRvL..99x1103B} to determine pseudoconstraints are computed only accounting for short periodic oscillations, typically at the order of magnitude of the mean motion of the Moon around the Earth. For instance, the recession motion of line of apsides in 8.85 years or the precession motion of the lunar orbit on the ecliptic plane in 18.6 years are both neglected. Therefore, this analytic solution remains only valid for a few years while LLR data span over 45 years (see also the discussion in footnote 2 from \cite{2015PhRvD..92f4049H}). In a more correct strategy, the SME modeling must be included in the complete data analysis and the SME coefficients need to be estimated in a global fit along with others parameters by taking into account short and long period terms and also correlations. This approach has recently been successfully used in a study using Very Long Baseline Interferometry data \cite{2016arXiv160401663L} to improve the estimation of the $\bar{s}^{T\!T}$ coefficient. In this Letter, we apply for the first time the same approach to estimate SME coefficients from LLR data.

\par LLR is used to conduct high-precision measurements of the light travel time of short laser pulses between a LLR station on Earth (McDonald Observatory in Texas, Observatoire de la C\^ote d'Azur in France, Haleakala Observatory in Hawaii, Apache Point Observatory in New Mexico, and Matera in Italy) to a corner cube retroreflector on the lunar surface (Apollo XI, XIV, XV and Lunokhod 1, 2) and back to the station receiver. The change of the round-trip travel time contains a lot of information about the Earth-Moon system leading to many different fields of investigations like lunar science, geodesy, geodynamics and gravitational physics. In addition, the determination of physical or gravitational parameters benefits from the 45 years of LLR data span and from the technology improvement that has led to the current observational accuracy at the subcentimetric level \cite{1998A&AS..130..235S,2013RPPh...76g6901M}. LLR data are presented as normal points that combine time series of measured light travel time of photons, averaged over several minutes to achieve a higher signal-to-noise ratio measurement of the lunar range at some characteristic epoch. Each normal point is characterized by one emission time, one time delay and some additional observational parameters such as laser wavelength, atmospheric temperature and pressure etc. According to \cite{1999A&A...343..624C}, the theoretical expression of the time delay is defined as
\begin{equation}
  \Delta{}t_c=\big[T_3-\Delta{}\tau_t(T_3)\big]-\big[T_1-\Delta{}\tau_t(T_1)\big]\text{,}
  \label{eq2}
\end{equation}
with
\begin{subequations}\label{eq2_b}
\begin{align}
  T_3&=T_2+\frac{\Vert\bm{r}_{o'}(T_3)\!-\!\bm{r}_r(T_2)\Vert}{c}+\Delta{}\tau_s+\Delta{}\tau_a\label{eq2_1}\\
  T_2&=T_1+\frac{\Vert\bm{r}_r(T_2)\!-\!\bm{r}_o(T_1)\Vert}{c}+\Delta{}\tau_s+\Delta{}\tau_a\text{.}\label{eq2_2}
\end{align}
\end{subequations}
$\Delta{}t_c$ is the theoretical round-trip travel time in international atomic time (TAI), $T_1$, $T_2$, and $T_3$ are respectively the barycentric dynamical time (TDB) at the emission, reflection, and reception points,  $\bm{r}_o$ and $\bm{r}_{o'}$ are respectively the barycentric position vector at the emitter and the reception point on Earth, $\bm{r}_r$ is the barycentric position vector of one of the five lunar retroreflectors, $\Delta{}\tau_s$ is the one-way gravitational time delay correction (i.e. the Shapiro time delay), $\Delta{}\tau_a$ is the one-way tropospheric correction to the light propagation and $\Delta{}\tau_t$ is a relativistic time scale correction due to the transformation between TDB and TAI (see \cite{2003AJ....126.2687S} for further details).

\par In order to analyze LLR data in the SME framework, we have built a new numerical lunar ephemeris, \'eph\'em\'eride lunaire parisienne num\'erique (ELPN) which computes numerically orbital and rotational motion of the Moon. In addition, ELPN computes the angular velocity of the Moon's liquid core considering a laminar damping term between the core and the lunar mantle since tidal and core dissipations present separable signatures as discussed in \cite{2001JGR...10627933W}.

\par As a validation of our dynamical model, we compare our GR solution with the DE430 solution from JPL \cite{2014IPNPR.196C...1F}. The main differences are the Moon gravitational potential (modeled until the fifth degree in ELPN versus sixth degree in DE430) and the number of accounted asteroids (70 in ELPN versus 343 in DE430). Moreover, we integrate the partial derivatives of the observables with respect to all the estimated parameters by including directly the variational equations in the integration (representing a total of 6000 integrated equations) instead of the pure numerical computation method implemented in DE430. The most important specificity of ELPN is the Lorentz violating contributions arising from the Earth-Moon system, implemented with the associated partial derivatives. The additional acceleration of the Earth-Moon vector due to SME is given in \cite{2006PhRvD..74d5001B} [see Eq. (104)] and is expressed as
\begin{align}
  a^J_{\text{LV}}=\ &\frac{G_{\!N}M}{r^3}\Big[\bar{s}^{J\!K}_tr^K\!-\frac{3}{2}\bar{s}^{K\!L}_t\hat{r}^K\hat{r}^Lr^J+3\bar{s}^{T\!K}\hat{V}^Kr^J\Big.\nonumber\\
  &-\Big.\bar{s}^{T\!J}\hat{V}^Kr^K-\bar{s}^{T\!K}\hat{V}^Jr^K+3\bar{s}^{T\!L}\hat{V}^K\hat{r}^K\hat{r}^Lr^J\Big.\nonumber\\
  &+\Big.2\frac{\delta{}m}{M}\Big(\bar{s}^{T\!K}\hat{v}^Kr^J-\bar{s}^{T\!J}\hat{v}^Kr^K\Big)\Big]\text{,}\label{eq3}
\end{align}
where $G_{\!N}$ is the observed Newtonian constant, $M$ is the mass of the Earth-Moon barycenter, and $\delta{}m$ is the difference between Earth and the lunar masses, $\hat{r}^J$ being the unit position vector of the Moon with respect to Earth, $\hat{v}^J\!=\!v^J\!/c$ with $v^J$ being the relative velocity vector of the Moon with respect to Earth, and $\hat{V}^J\!=\!V^J\!/c$ with $V^J$ being the heliocentric velocity vector of the Earth-Moon barycenter. Latin indices are used to denote space coordinate ($X,Y,Z$) and $T$ represents the time coordinate (TDB) as in Eq. \eqref{eq2} (see also \cite{2006PhRvD..74d5001B} for the conventions used in SME analyses). In the last equation, we used the three-dimensional traceless tensor $\bar{s}^{J\!K}_t=\bar{s}^{J\!K}\!-\!\frac{1}{3}\bar{s}^{T\!T}\delta^{J\!K}$ and a rescaled observable Newtonian constant defined as $G_{\!N}=G(1+\frac{5}{3}\bar{s}^{T\!T})$ \cite{2013PhRvD..88j2001B}. 

\par The numerical ephemeris provides the position, velocity and orientation of the different bodies and all the associated partial derivatives. The remaining quantities needed for the evaluation of Eq. \eqref{eq2} are computed using an existing software at the Paris Observatory Lunar Analysis Centre (POLAC) based on the 2010 international Earth rotation system conventions \cite{2010ITN....36....1P}. This software has been upgraded in order to take into account effects from the breaking of the Lorentz symmetry on the light propagation. More precisely, the SME time delay formula [see Eq. (24) from \cite{2009PhRvD..80d4004B}] of the pure gravity sector expressed in standard harmonic gauge has been used for the computation of $\Delta\tau_s$ in Eqs. \eqref{eq2_b}. This expression has to take into account the rescaled Newtonian constant $G_{\!N}$ defined previously. The SME gravitational time delay is taken into account for consistency since it is unobservable considering the smallness of SME coefficients and the current accuracy of LLR observations. We finally determine residuals using LLR data and minimize them with a standard iterative least-square fit. 

\par First of all, we built a reference solution computed in pure GR by adjusting a set of 76 parameters including the geocentric positions of LLR stations, the selenocentric positions of lunar retroreflectors, the barycentric Earth-Moon position and velocity vectors at J2000, the lunar libration angles with their time derivatives at J2000 and the rotation vector of the Moon fluid core at J2000. We also estimated the masses of the Earth-Moon system, the Earth rotational time lag for diurnal and semidiurnal deformation, the potential Love number of degrees 2, 3, and 4 of the Moon, the Moon time lag for solid-body tide of degree 2, the total moment of inertia of the Moon, the ratio of polar moment of inertia of core to the mean total moment of inertia of the Moon, the flattening of the Moon core and the damping term between the solid mantle and the fluid core of the Moon. After this fitting process, the differences between the ELPN solution and DE430 remain below 5 cm on the Earth-Moon distance and below 50 cm along the lunar orbit during the time span of LLR data meaning that the two ephemerides are very similar. Moreover, the differences between our estimated values for the different parameters and the estimated values in DE430 remain below the 5-sigma uncertainty.

\par This new lunar solution constituted the starting point of the analysis that includes Lorentz violation terms. From it, we built a second solution by adjusting the exact same parameters together with the SME coefficients. This analysis reveals that two pairs of SME coefficients are highly correlated (i.e. the absolute value of their correlation coefficient is higher than 0.99). This indicates that the data are sensitive to linear combinations of these parameters only. An analysis of the partial derivatives and of the covariance matrix allows us to determine the linear combinations to which the data are sensitive to:
\begin{equation}
  \begin{array}{r l r l}
  	\bar{s}^{A}&\!\!=\bar{s}^{X\!X}-\bar{s}^{Y\!Y}      &\ \ \ \bar{s}^{B}&\!\!=\bar{s}^{X\!X}+\bar{s}^{Y\!Y}-2\bar{s}^{Z\!Z}\\
  	\bar{s}^{C}&\!\!=\bar{s}^{T\!Y}+0.43\bar{s}^{T\!Z}  &\ \ \ \bar{s}^{D}&\!\!=\bar{s}^{B}-4.5\bar s^{Y\!Z}\text{.}\label{eq:lin}
  \end{array}
\end{equation}
These linear combinations have to be compared to those of \cite{2015PhRvD..92f4049H} [cf., Eqs. (16)] where authors used Eqs. (107) in \cite{2006PhRvD..74d5001B}. We notice that linear combinations involving $\bar{s}^{T\!J}$ are similar [see Eq (16c) and (16d)], meaning that oscillating signatures derived in \cite{2006PhRvD..74d5001B} are well determined for $\bar{s}^{T\!J}$. However, the other linear combination is different. Note that in all SME analyses, $\bar{s}^A$ and $\bar{s}^B$ are used instead of $\bar{s}^{X\!X}$, $\bar{s}^{Y\!Y}$, and $\bar{s}^{Z\!Z}$. These two combinations enforce the traceless condition on $\bar{s}^{\mu\nu}$ (see also \cite{2006PhRvD..74d5001B}). A new adjustment using these linear combinations of SME coefficients provides estimations and statistical uncertainties on the different linear combinations of the SME coefficients: $\bar{s}^{T\!X}$, $\bar{s}^{X\!Y}$,  $\bar{s}^{X\!Z}$, $\bar{s}^{A}$, $\bar{s}^{C}$ and $\bar{s}^{D}$. Moreover, in this new solution, the six fitted linear combinations do not show high correlations (below 50\% between $\bar{s}^{T\!X}$ and $\bar{s}^{A}$).

\par As mentioned by \cite{1996PhRvD..53.6730W}, we expect LLR data analysis to suffer from systematic uncertainties in model parameter estimates. Such systematics may arise from observations or from mismodeling, for instance, from neglected correlation between observations of each LLR station. As a consequence, the standard deviation reported by the least-square fit (called the statistical uncertainty labeled $\sigma_{\text{stat}}$) underestimates realistic uncertainty. Therefore, it is essential to quantify the order of magnitude of such systematics in the data analysis. 

\par
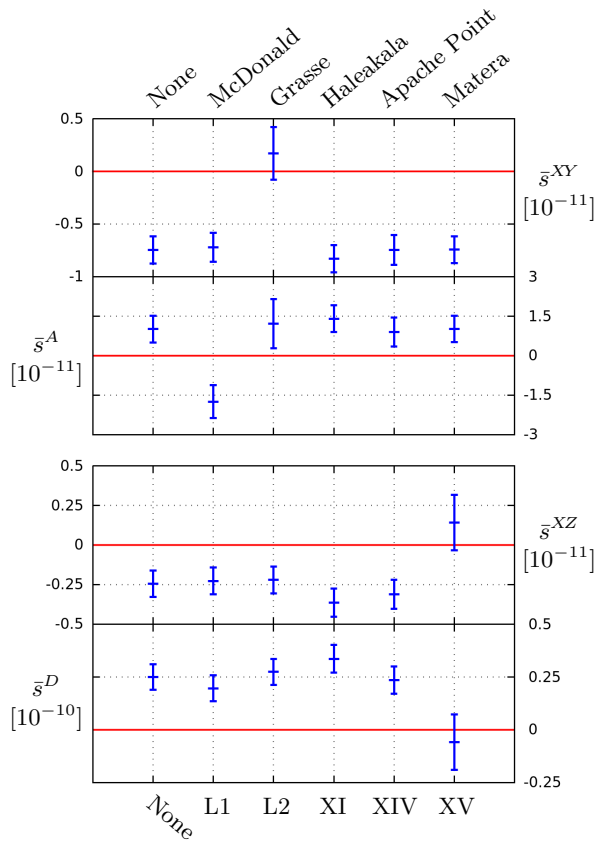
\begin{figure} 
  \begin{center}
    \includegraphics[width=7cm, clip=true, trim=0cm 2cm 0cm -5cm]{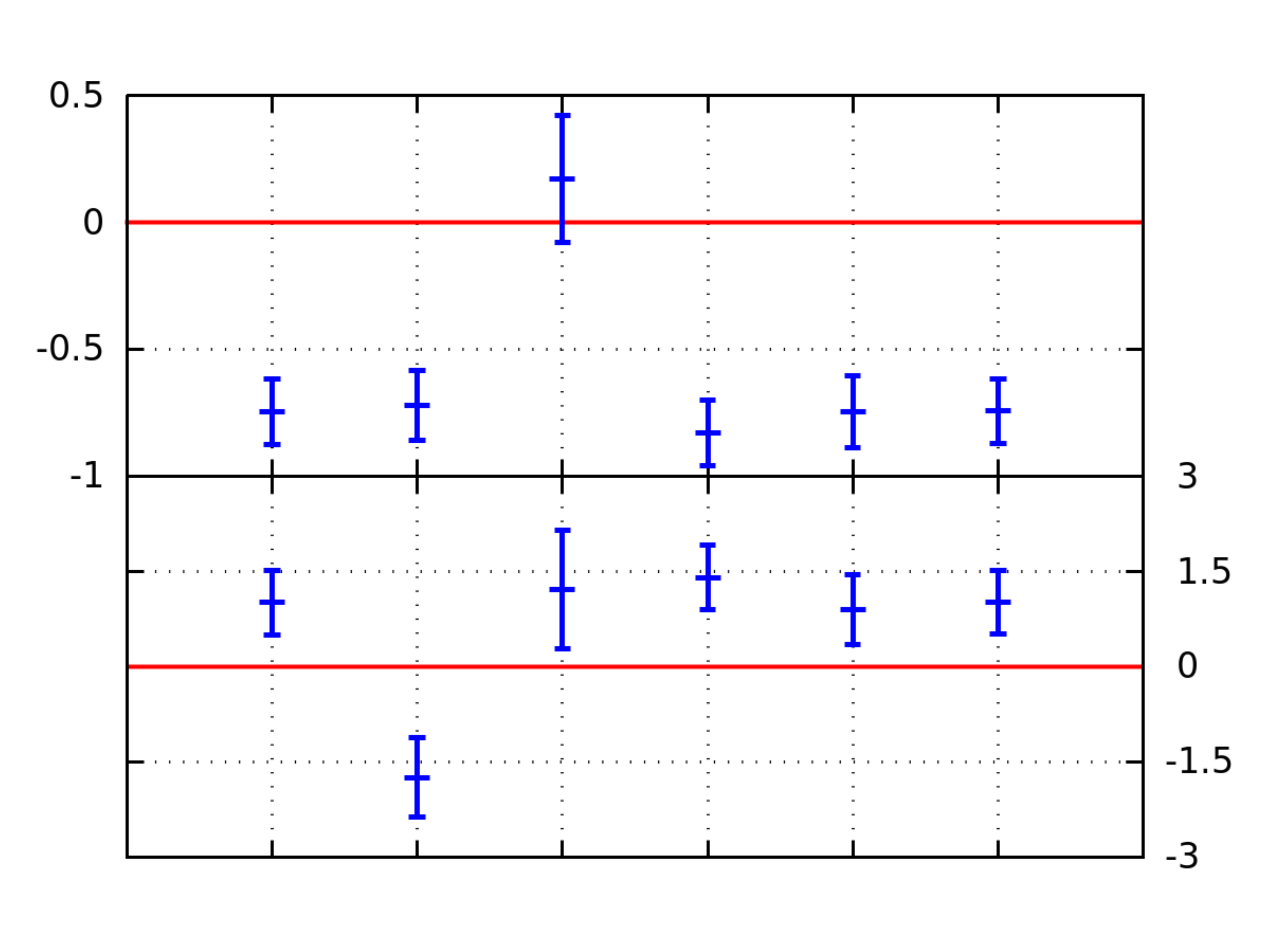}
	\includegraphics[width=7cm, clip=true, trim=0cm 1cm 0cm 2cm]{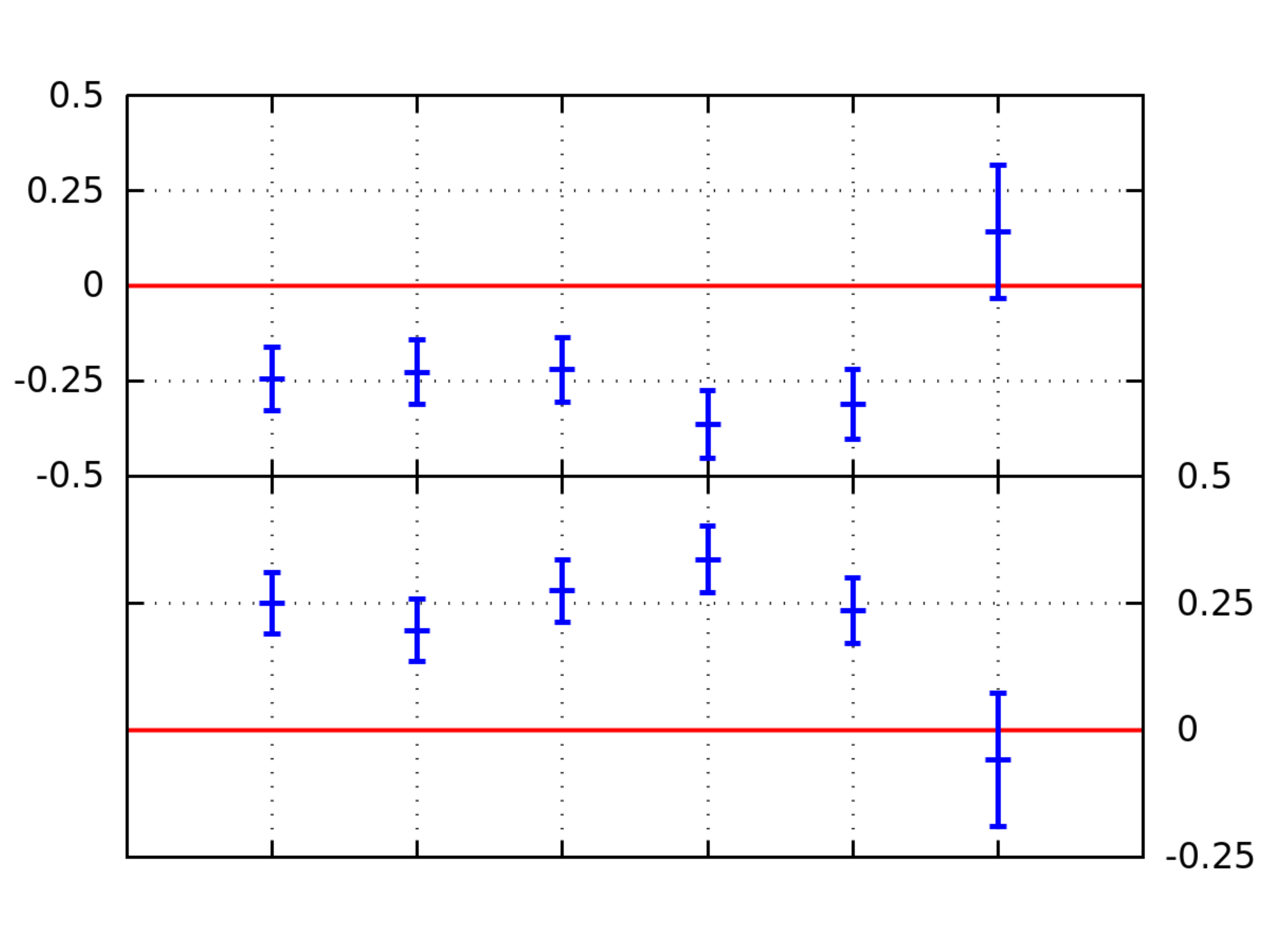}
  \end{center}
  \setlength{\unitlength}{1.0769cm}
  \begin{picture}(1,1)
    \put(3.25,4.35){\shortstack[c]{$\bar{s}^{X\!Z}$\\$[10^{-11}]$}}
	\put(-3.08,2.4){\shortstack[c]{$\bar{s}^{D}$\\$[10^{-10}]$}}
	\put(-1.4,1.4){\rotatebox{-40}{$\text{None}$}}
	\put(-0.68,1.3){\rotatebox{0}{$\text{L1}$}}
	\put(0.05,1.3){\rotatebox{0}{$\text{L2}$}}	
	\put(0.75,1.3){\rotatebox{0}{$\text{XI}$}}	
	\put(1.4,1.3){\rotatebox{0}{$\text{XIV}$}}	
	\put(2.22,1.3){\rotatebox{0}{$\text{XV}$}}
	\put(3.25,8.7){\shortstack[c]{$\bar{s}^{X\!Y}$\\$[10^{-11}]$}}
	\put(-3.08,6.65){\shortstack[c]{$\bar{s}^{A}$\\$[10^{-11}]$}}
	\put(-1.4,10){\rotatebox{40}{$\text{None}$}}
	\put(-0.65,10){\rotatebox{40}{$\text{McDonald}$}}
	\put(0.1,10){\rotatebox{40}{$\text{Grasse}$}}
	\put(0.75,10){\rotatebox{40}{$\text{Haleakala}$}}
	\put(1.45,10){\rotatebox{40}{$\text{Apache Point}$}}
	\put(2.25,10){\rotatebox{40}{$\text{Matera}$}}
  \end{picture}
  \vspace{-1.25cm}
  \caption{Top: estimations of $\bar{s}^{X\!Y}$ and $\bar{s}^{A}$ as a function of data subsamples by LLR stations. Each station name along $x$ axis corresponds to the subsample without data from the corresponding station. Bottom: estimations of $\bar{s}^{X\!Z}$ and $\bar{s}^{D}$ as a function of data subsamples by lunar reflectors. L1 and L2 correspond to subsamples without respectively Lunokhod 1 and 2 data, while XI, XIV and XV refer to subsamples without respectively Apollo XI, XIV, and XV data. The top and bottom error bars are those provided by the chi-square fit at $1\sigma_{\text{stat}}$ standard deviation and the red line corresponds to the theoretical values of the SME coefficients in the GR framework.}
	\label{fig_systematic}
\end{figure}

 In order to assess the impact of potential systematics, we split our data set into five independent subsets by removing data related to one of the five LLR stations. We estimated the parameters mentioned above with these five subsets. The top of Fig. \ref{fig_systematic} shows for illustrative purposes the derived estimations on two SME coefficients. We can see that the confidence intervals derived with the different stations do not overlap. The SME coefficients being universal, this result is a strong indication of the presence of systematics. A similar analysis has been performed with the different lunar reflectors and is presented for illustrative purposes on the bottom of Fig. \ref{fig_systematic}. For each subsamples we have checked that the new estimations of all parameters (SME coefficient and others) stay below the $5\sigma_{\text{stat}}$ confidence interval of the reference solution, meaning that the new solution is valid.

\par In order to estimate systematic uncertainties, we used a jackknife resampling method \cite{1993stp..book.....L,Gottlieb2003} (see a similar use of this resampling technique in the context of asteroid observations \cite{2009Icar..204..145M} or in the context of cosmology \cite{2008MNRAS.389..766P,2009ApJ...691.1307H}). The idea is to split the data set into $n$ different independent subsets and to estimate the parameters by systematically excluding one of the subset. If we denote by $x_i$ the $n$ estimations of a parameter of interest $x$ obtained by removing one subset of data, an estimate of the systematic variance is $\sigma^2(x)=\frac{n-1}{n}\sum_i^n(x_i-\bar{x})^2$, with $\bar{x}$ being the mean of the $n$ values $x_i$ \cite{1993stp..book.....L}. We applied this resampling method to the estimates of the SME parameters for two cases: (i) by splitting our data set with respect to the different LLR stations (the obtained systematic variance is denoted $\sigma_s^2$) and (ii) by splitting our data set with respect to the different lunar reflectors (the obtained systematic variance is denoted $\sigma_r^2$). The total variance estimate is the sum of the statistical and of the two estimated uncertainties obtained with the resampling method, $\sigma^2=\sigma_{\text{stat}}^2+\sigma_s^2+\sigma_r^2$. Our estimations of the SME coefficients and their realistic errors are reported in Table~\ref{tab1}.

\begin{table} 
  \begin{tabular}{c|c|c}
    \toprule
	SME              & Other works & This work \\
	\hline
	$\bar{s}^{T\!X}$ & $(+5.2\pm5.3)\times10^{-9\ }$  & $(-0.9\pm1.0)\times10^{-8\ }$ \\
	$\bar{s}^{X\!Y}$ & $(-3.5\pm3.6)\times10^{-11}$   & $(-5.7\pm7.7)\times10^{-12}$  \\
	$\bar{s}^{X\!Z}$ & $(-2.0\pm2.0)\times10^{-11}$   & $(-2.2\pm5.9)\times10^{-12}$  \\
	$\bar{s}^A$      & $(-1.0\pm1.0)\times10^{-10}$   & $(+0.6\pm4.2)\times10^{-11}$  \\
	$\bar{s}^C$      & $(-1.0\pm0.9)\times10^{-8\ }$  & $(+6.2\pm7.9)\times10^{-9\ }$ \\
	$\bar s^D$      & $(-1.2\pm1.2)\times10^{-10}$  & $(+2.3\pm4.5)\times10^{-11}$ \\
	\botrule
  \end{tabular}
  \caption{Table of estimated values of SME parameters of the minimal SME with LLR data. Second column: results deduced from a postfit analysis of binary pulsars observations in \cite{2014PhRvL.112k1103S}. Third column: results from this work obtained performing a global fitting to LLR data. The quoted uncertainties correspond to $1\sigma$ realistic uncertainties based on the statistical and systematic errors. The linear combinations of the SME coefficients are defined in Eqs. (\ref{eq:lin}).}
  \label{tab1}
\end{table}

\par Some of our estimates improved previous constraints based on postfit analysis by a factor up to 5. More precisely, the constraints on the $\bar{s}^{T\!J}$ coefficients are of the same order of magnitude as the ones from binary pulsars \cite{2014PhRvL.112k1103S} but improve the ones from the planetary ephemerides by a factor 5 \cite{2015PhRvD..92f4049H}. The estimates on $\bar{s}^{X\!Y}$ and $\bar{s}^{X\!Z}$ improve previous constraints from binary pulsars by a factor $4-5$ and from planetary ephemerides by 1 order of magnitude. The estimates $\bar{s}^A$ and $\bar s^D$ are improved by a factor 2.5 with respect to binary pulsar analysis and by 1 order of magnitude with respect to planetary ephemerides. In addition our results improve a previous postfit analysis with LLR data \cite{2007PhRvL..99x1103B} by a factor 30 to 800. Nevertheless, we want to emphasize that the linear combinations fitted in that paper have been determined in a sensitivity analysis based on theoretical calculations (see \cite{2006PhRvD..74d5001B}) only accounting for short periodic oscillations (see discussion above). Our numerical analysis shows that this approach is not accurate enough for a full data analysis since the fitted linear combinations are different.

\par As mentioned above, our results are mainly dominated by systematic uncertainty. One way to improve our estimates would be to improve our understanding of these and to model them carefully. Moreover, some SME coefficients (mainly $\bar{s}^{A}$) show slight correlations with parameters appearing in the rotational motion of the Moon as the principal moment of inertia (at the level 0.85), the quadrupole moment (at the level 0.87), the potential Stockes coefficient $C_{22}$ (at the level 0.81) and the polar component of the velocity vector of the fluid core (at the level 0.85). Those parameters have an impact on the rotational motion of the Moon that affects the orbital motion through the effect of the lunar potential. Consequently, it would be interesting to produce a joint GRAIL \cite{2014GeoRL..41.1452K,2014GeoRL..41.3382L} and LLR data analysis. We expect this combined analysis to help in decorrelating the SME parameters from the lunar potential and therefore to improve marginalized estimations of the SME coefficients. 

\par In conclusion, we have analyzed a set of 20721 LLR data spanning 44 years by using ELPN, a new numerical lunar ephemeris. In this work, the SME modeling has been included in the complete data modeling and the coefficients of the minimal SME are estimated simultaneously with other LLR standard parameters. We show that the data are sensitive to linear combinations of the SME coefficients that have been determined numerically. We provided an estimation of these combinations taking into account statistical and systematic uncertainties. We found no evidence for Lorentz violation at the 1$\sigma$ confidence level. Our results improve several constraints on the SME coefficients with respect to previous studies \cite{2014PhRvL.112k1103S,2015PhRvD..92f4049H,2007PhRvL..99x1103B}. In addition previous studies are based on postfit analysis and therefore neglect all potential correlations (see also the discussion in \cite{2016arXiv160401663L}). For this reason, our estimates are more robust.

\par \emph{Acknowledgments} - The authors thank P. Wolf and Q. Bailey for useful comments on a preliminary version of this paper. They are also grateful to LLR staffs at C\^ote d'Azur, McDonald and Apache Point observatories for providing their observations. A. B. and C. L. P. L. are grateful for the financial support of CNRS/GRAM and "Axe Gphys" of Paris Observatory Scientific Council.
 
\bibliographystyle{apsrev4-1}
\bibliography{SME_LLR}

\end{document}